\documentclass[A4paper,twocolumn,aps,pra,showpacs,
fixfloats]{revtex4-1}
\usepackage{bm}
\usepackage{dcolumn}
\usepackage{graphicx}
\begin{document}

%
%\title[Multiple occurrences  of the breakdown...]{Multiple occurrences of the breakdown of the dipole and dipole-quadrupole approximation in spectra of fullerene anions, C$_{n}^{-}$:
%the initial insight}
%
\title{On possible multiple occurrences of the breakdown of the dipole and dipole-quadrupole approximations in angle-differential spectra of
 fullerene anions, C$_{n}^{-}$}
\author{ V K Dolmatov}
\email[Send e-mail to:\ ]{vkdolmatov@una.edu}
\author{A Edwards}
\email{aedwards1@una.edu}
\author{C G Lane}
\email{cglane@una.edu}
\altaffiliation{Presently with Brentwood Services, Inc., Brentwood, TN 37027}
\address{Department of Physics and Earth Science, University of North Alabama,
Florence, Alabama 35632, USA}
%
%\ead{vkdolmatov@una.edu}
\begin{abstract}
We provide the initial insight into the angle-differential photodetachment spectra of fullerene anions beyond the dipole approximation by utilizing a broadly used modelling of C$_{n}$. In the model, the C$_{n}$ cage is approximated by a spherical attractive potential of a certain inner radius, thickness and depth which binds an external electron, thereby turning into a C$_{n}^{-}$ anion. It is demonstrated in the framework of the utilized model, which a single-electron model in its essence, that the dipole and dipole-quadrupole approximations might get broken down in the angle-differential photodetachment spectra of fullerene anions, C$_{n}^{-}$, at a great number of photon energies of only a few tens of eV. Moreover, the breakdown occurrences are shown to start developing at lower photon energies, and the frequency of the occurrences grows with increasing size of C$_{n}^{-}$. The findings are demonstrated by direct calculations of angle-differential photodetachment cross sections of the C$_{60}^{-}$, C$_{240}^{-}$, C$_{540}^{-}$ and C$_{1500}^{-}$ anions. It is not clear beforehand how electron correlation might affect the predicted anomalies in the C$_{n}^{-}$ angle-differential photodetachment spectra, so that the verification(s) of the predicted breakdowns by means of more sophisticated theories is urged.

\end{abstract}

%\pacs{32.80.Fb, 32.80.Dz}
%\submitto{\jpb}
\maketitle
\section{Introduction}
One of the spectacular breakthroughs in atomic physics in the recent  two to two and a half decades  has been the discovery of the insufficient use (the breakdown) of a dipole approximation alone to understand adequately the angle-differential photoionization cross sections, $\frac{d\sigma}{d\Omega}$, of atoms and molecules at photon energies ranging from a few keV to, surprisingly, as low as only a few tens of eV \cite{HemmersRPC04,KrassigRPC04,AmusiaRPC04} (and references in these review papers). It was shown that
a second term in the Taylor series expansion $\Delta V \propto {\bf A}_{\omega}({\bf r},t)\hat{\bf p} \propto  [\hat{\bf e}\hat{\bf p} + i({\bf k\hat{r}})(\hat{\bf e}\hat{\bf p})]{\rm e}^{-i\omega t}$ of the
photon-electron interaction ($\Delta V$) cannot be ignored in the calculation of  $\frac{d\sigma}{d\Omega}$, at those energies. Here, the first term in the brackets is referred to as the electric-dipole term ($E1$),
the second term is the electric-quadrupole term ($E2$), ${\bf \hat{r}}$ and
$\hat{\bf p}$ are the electron's position-vector and momentum operators, respectively, ${\bf k}$ and $\omega$ are the photon momentum and energy (in atomic units), respectively, and $\hat{\bf e}$ is  the unit vector of the photon polarization. The cross-product between the $E1$ and $E2$ terms, referred to as the  $E1$-$E2$ interference  term, vanishes
in the calculation of the total angle-integrated cross section, but holds and plays an important role in the calculation of $\frac{d\sigma}{d\Omega}$.

Relatively recently, the study of the $E1$-$E2$ effects has been extended to double-electron photoionization of atoms
\cite{ManakovJPB06},  two-photon photoionization of atoms \cite{GrumJPB12}, photoionization of endohedral fullerenes/atoms $A$@C$_{60}$ \cite{AmusiaPRA04,DolmatovPRA08,DeshmukhJPB09}, including
photoionization by strong femtosecond VUV pulses \cite{GrumJPB11},
multiwalled onions
$A$@C$_{n}$@C$_{m > n}@...$ \cite{DolmatovPRA08,AmusiaPRA09} and $A$@C$_{60}^{\pm z}$ ions \cite{DolmatovPRA06,DolmatovPRA09}.
It has been shown that their nondipole angle-differential photoionization spectra develop significant resonances versus the photon energy. The nature of these resonances is the same as the nature of resonances in dipole photoionization of $A$@C$_{n}$ atoms \cite{Baltenkov,CRs} (also refer to works cited above and a review paper \cite{DolmAQC} on the subject with numerous references therein). These resonances are termed the confinement resonances \cite{CRs}. They are
due to interference of the photoelectron waves emerging directly from the confined atom $A$ and those scattered off the C$_{n}$ cage. The predicted existence of confinement resonances has relatively recently been
confirmed experimentally \cite{Phaneuf1,Phaneuf2}.

Confinement resonances have also been theoretically predicted in photodetachment spectra, both dipole total and angle-differential, of fullerene anions \cite{Lohr92,AmBaltKrak}. In work \cite{Lohr92}, such predictions were made
 in the framework of a model for the fullerene anion C$_{60}^{-}$ in which the attached electron was regarded to be moving in an attractive spherical Dirac-$\delta$-function potential $U(r)=-A\delta(r-R_{0})$ while residing in a state with a given orbital quantum number $\ell$. In the thus defined C$_{60}$ potential, $R_{0}$ is the mean radius of of C$_{60}$, and the constant $A$ is to be found by matching the model electron affinity (EA) of C$_{60}^{-}$ to its experimental value. The authors of \cite{Lohr92} calculated photodetachment cross sections of C$_{60}^{-}$
both from the $s$-like state and the $p$-like state of the attached electron and discovered the existence of resonant maxima and minima in the dipole photodetachment cross section of, and dipole photoelectron angular distribution from, C$_{60}^{-}$ due to a standing photoelectron wave formed inside the hollow interior of C$_{60}$. The same observation was later reported in work \cite{AmBaltKrak}, where the same modelling of C$_{60}$ by the same
 attractive spherical Dirac-$\delta$-function potential $U(r)=-A\delta(r-R_{0})$, as in work \cite{Lohr92}, was employed to study photodetachment of, and elastic electron scattering off, C$_{60}^{-}$.
The predicted resonances in the photodetachment cross section of C$_{60}^{-}$ are, thus, of the same origin as the above discussed confinement resonances in the total and differential photoionization cross sections of endohedral atoms, $A$@C$_{60}$. It is, however, both interesting and important to note that the confinement resonant minima in the dipole photodetachment cross sections of C$_{60}^{-}$ are very deep; there, the dipole photoionization amplitudes take small values. This is in contrast to the confinement resonant minima in the photoionization cross sections of $A$@C$_{60}$ where the cross section graph oscillates about the photoionization cross section graph for a free atom, thereby preventing the cross section to drop to a very low value, see, e.g., \cite{AmusiaPRA04}.

To the best of the authors' knowledge, nothing is really known to date about the specificity of the electric dipole-quadrupole effects in regard to angle-differential photodetachment spectra of  C$_{n}^{-}$ anions. Meanwhile, such study is
particularly interesting. This is because a fullerene anion is, in essence, a giant atom since the attached electron belongs to the ${\rm C}_{n}$ cage that has a large radius. Indeed, the number of the carbon atoms in a C$_{n}$ cage can be in the limits of thousands of atoms \cite{GiantCns}. This results in a giant size of the fullerene. For example, the radius of the C$_{3380}$ is more than $50 a_{0}$ ($a_{0}$ being the first Bohr radius of the hydrogen atom). One can, thus, reasonably expect a much greater significance of the quadrupole term $({\bf kr})\hat{\bf e}\hat{\bf p}$  in the  interaction of a photon with the attached electron of C$_{n}^{-}$, compared to the photon interaction  both with an isolated atom $A$ or atom $A$ encapsulated inside C$_{n}$, $A$@C$_{n}$. This is because a much larger size of C$_{n}^{-}$ versus
the size of an atom, be it a free atom or atom encapsulated inside C$_{n}$, can compensate, to some degree, for a low value of the photon momentum. Furthermore, it is both interesting and important to evaluate the validity of the dipole ($E1$) and dipole-quadrupole $E1 - E2$ approximations in the angle-differential photodetachment cross sections in view of deep confinement-resonance-minima in the dipole photodetachment amplitudes. It is not clear beforehand to what degree the dipole and dipole-quadruple approximations might survive or break down at and/or near these minima in photodetachment spectra of C$_{n}^{-}$. The situation here differs considerably from a similar situation in $A$@C$_{60}$ photoionization \cite{AmusiaPRA04} where said minima in the dipole photoionization amplitudes are noticeably shallower than in the case of fullerene anions. In the former case, thus, naturally, the overall role of the nondipole
 terms in the angle-differential photoionization cross section of $A$@C$_{60}$ is lessened, although it cannot be ignored in corresponding studies. Note, it is not the first time observation of how the presence of an encapsulated atom inside C$_{60}$ may change the situation radically compared to empty C$_{60}$. Indeed, for example, as was originally unraveled in \cite{DolmatovJPB14} and later demonstrated in \cite{DolmatovPRA15a,DolmatovPRA15} as well, the cross section of elastic electron scattering off $A$@C$_{60}$ may be smaller than off empty C$_{60}$, or, what was the mostly surprising finding, off the isolated atom $A$ itself, at some energies.

It is precisely the aim of the present work to get insight into the specificity of $E1$-$E2$ effects in photoelectron angular distributions from fullerene anions, including their evolution with the
increasing size of the anion. The present study focuses on the calculation of said effects upon photodetachment of  C$_{n}^{-}$s with $n = 60$, $240$, $540$ and $1500$.
at photon energies of only a few tens eV.

In contrast to \cite{Lohr92,AmBaltKrak}, the study is carried out in a model approximation,
where the C$_{n}$ cage is modelled by an attractive spherical potential of certain depth $U_{0}$, inner radius $r_{0}$ and thickness $\Delta$, as in many earlier \cite{PushkaPRA93,XuPRL96}  and recent
 \cite{AmusiaPRA04,DolmatovPRA08,DeshmukhJPB09,GrumJPB11,DolmatovPRA06,DolmatovPRA09,Winstead,GorczycaPRA12,DolmatovJPB14,DolmatovPRA15a,DolmatovPRA15} studies of the  spectra of $A$@C$_{n}$'s. Thus, a fullerene anion itself is due to the binding of an external electron by this potential into a $n\ell$-state. In the present work, similar to work \cite{Lohr92}, we perform calculations for both the $s$-like and $p$-like states of the attached electron.

It is shown in the present work that  not only the electric dipole approximation, but also the electric dipole-quadrupole approximation breaks down in the angular differential photodetachment spectra of fullerene anions at a great number of points over the photon energy. Furthermore, this number is found to increase, whereas corresponding critical photon energies shift toward a lower end of the spectrum, with increasing size of a fullerene anion.

Since it is not obvious beforehand whether the account for electron correlation, omitted in the present study,  will or will not `help' the dipole and dipole-quadrupole approximations to `overcome' the predicted breakdown,
a more sophisticated study, than the one performed in the present work, would be of a great interest. Such study, however, is beyond the capability of the present authors at the present time. Rather, we believe that the results of
the present study will encourage other researchers to perform experimental or more sophisticated theoretical studies of the angle-differential photodetachment cross sections of fullerene anions, now that the present work has
highlighted the most intriguing parts of the problem that are in need of a deeper study.

Atomic units (\textit{au}) are used throughout the text unless specified otherwise.

\section{Theoretical concepts}
\subsection{Electric dipole-quadrupole $E1$-$E2$ inteference}
With account for $E1$-$E2$ electric dipole-quadrupole  interference, the angle-differential photoionization cross section, $\frac{d\sigma_{n\ell}}{d\Omega}$, of a $n\ell$-subshell  of a spherical target by a linearly polarized light reads \cite{CooperPRA93}:
\begin{eqnarray}
\frac{d \sigma_{n\ell}}{d \Omega}=
\frac{\sigma_{n\ell}}{4 \pi} \left\{\left[
1 + \frac{\beta_{nl}}{2}(3 \cos^{2}\theta -1)\right]+ \Delta E_{12}\right\}.
\label{eqsigma}
\end{eqnarray}
Here, $\sigma_{n\ell}$ is the total dipole photoionization cross section of a $n\ell$-subshell of the target,
the term in brackets represents the $E1$-$E1$ electric dipole  interference term, where  $\beta_{n\ell}$ is called the dipole photoelectron angular-asymmetry
parameter, whereas $\Delta E_{12}$ represents the $E1$-$E2$ electric dipole-quadrupole  interference term:
\begin{eqnarray}
\Delta E_{12}= (\delta_{n\ell}+
\gamma_{n\ell}\cos^{2}\theta)\sin\theta \cos\phi.
\label{eqE_{12}}
\end{eqnarray}
Here, the direction of the photoelectron momentum, {\bf p}, is specified by the spherical angles $\theta$ and $\phi$ in the $XYZ$-system
of coordinates (the $Z$-axis points in the direction of the photon polarization vector $\hat{{\bf e}}$, the $X$-axis points in the direction of the photon momentum {\bf k}),
$\gamma_{n\ell}$ and $\delta_{n\ell}$ are the dipole-quadrupole (nondipole) photoelectron angular-asymmetry parameters. The parameters $\beta_{n\ell}$,
$\gamma_{n\ell}$ and $\delta_{n\ell}$ are defined as follows \cite{CooperPRA93}:
\begin{eqnarray}
\beta_{n\ell} = &&
\frac{
\ell (\ell-1) d^{2}_{\ell-1}+(\ell+1)(\ell+2) d^{2}_{\ell+1}
}
{
(2\ell+1)[l d^{2}_{\ell-1} + (\ell+1) d^{2}_{\ell+1}]
} \nonumber \\
&&
- \frac{
6\ell(\ell+1)d_{\ell-1}d_{\ell+1}\cos(\delta_{\ell+1}-\delta_{\ell-1})
}
{
(2\ell+1)[\ell d^{2}_{\ell-1} + (\ell+1) d^{2}_{\ell+1}]
},\\
\gamma_{n\ell}=&&
\frac{3 k}{2[\ell d^{2}_{\ell-1}+(\ell+1)d^{2}_{\ell+1}]} \nonumber \\
&&
\times \sum\limits_{\ell',\ell''}A_{\ell',\ell''}d_{\ell'}q_{\ell''}\cos(\delta_{\ell''}-
\delta_{\ell'}),
%\nonumber \\
\\
\delta_{n\ell}=&&
\frac{3 k}{2[\ell d^{2}_{\ell-1}+(\ell+1)d^{2}_{\ell+1}]} \nonumber \\
&&
\times \sum\limits_{\ell',\ell''}B_{\ell',\ell''}d_{\ell'}q_{\ell''}\cos(\delta_{\ell''}-
\delta_{\ell'}).
\label{eqgamma}
\end{eqnarray}
Here, $d_{\ell'}$ and $q_{\ell''}$ are the radial dipole and quadrupole
photoionization matrix elements, respectively:

\begin{eqnarray}
d_{\ell'}=\int_{0}^{\infty}{P_{\epsilon \ell'}(r) r P_{n\ell}(r) dr}, \\
q_{\ell''}=\int_{0}^{\infty}{P_{\epsilon \ell''}(r) r^{2} P_{n\ell}(r) dr}.
\label{eqdq}
\end{eqnarray}

In the  equations above, the wavefunctions $P_{n\ell}(r)$ and $P_{\epsilon \lambda}(r)$ ($\lambda \equiv \ell'$ or $\ell''$) are the solutions of the radial Schr\"{o}dinger equations for
the bound ($n\ell$) and scattering ($\epsilon \lambda$) states, respectively,
$\ell'= \ell \pm1$, $\ell''=\ell, \ell \pm 2$,
$\delta_{\lambda}$ is the phase of the wavefunction of a photoelectron, the
coefficients
$A_{\ell',\ell''}$ and $B_{\ell',\ell''}$ are presented in \cite{CooperPRA93}.

For the analyzer located at the magic angle $\theta = 54.7^{\rm o}$ in the
$\phi=0^{\rm o}$ plane, which is the geometry normally used in experiments on $E1$-$E2$ interference \cite{HemmersRPC04},
equation (\ref{eqsigma}) takes a particularly simple form:
\begin{eqnarray}
\frac{d \sigma_{n\ell}}{d \Omega}=\frac{\sigma_{n\ell}}{4 \pi}\left[1+\sqrt{\frac{2}{27}}\zeta_{n\ell}\right], \quad \zeta_{n\ell}=\gamma_{n\ell}+3\delta_{n\ell}.
\label{magic}
\end{eqnarray}

For photodetachment of a ${\rm s}$-electron the $\zeta_{\rm s}$ parameter takes the simplest form.
\begin{eqnarray}
\zeta_{\rm s} = 3\kappa\frac{q_{2}}{d_{1}}\cos(\delta_{2}-\delta_{1}),
\label{zetas}
\end{eqnarray}
%
%\begin{eqnarray}
%\zeta_{\rm p}= &&\frac{3k}{5(d_{0}^{2}+2d_{2}^{2})}
%\left[(7q_{3}\cos\delta_{32}-3q_{1}\cos\delta_{12})d_{2}\right. \nonumber \\
%&&\left. + (3q_{1}\cos\delta_{10}-2q_{3}\cos\delta_{30})d_{0}\right],
%\label{zetap}
%\end{eqnarray}
%
%
For for the case of a ${\rm p}$-electron the expression for $\zeta_{\rm p}$ is somewhat more complicated:
\begin{eqnarray}
\zeta_{\rm p}= &&\frac{3k}{5}
\left[\frac{(7q_{3}\cos\delta_{32}-3q_{1}\cos\delta_{12})d_{2}}{d_{0}^{2}+2d_{2}^{2}}  \right. \nonumber \\
&&\left. + \frac{(3q_{1}\cos\delta_{10}-2q_{3}\cos\delta_{30})d_{0}}{d_{0}^{2}+2d_{2}^{2}}    \right].
\label{zetap}
\end{eqnarray}
Here, $\delta_{\ell'\ell''}\equiv \delta_{\ell''}-\delta_{\ell'}$.

In the aims of the present paper, the authors focus on studying and presenting the results for the calculated $(d\sigma_{n\ell}/d\Omega)/(\sigma_{n\ell}/4\pi)$ ratio  at the magic angle, in
accordance with (\ref{magic}). The authors opt for presenting the results for the ratio only, because it carries all the needed information about the strength of the $E1$-$E2$ dipole-quadrupole effects
in relation to the dipole effects in the most illustrative and concise way. Indeed, the first term, equalled to unity, is associated with the dipole approximation, the second one is associated with the $E1$-$E2$ approximation, and it is easy to determine the $\zeta_{n\ell}$ parameter itself from the ratio, if needed.

\subsection{Modelling of a fullerene anion }

In the present work, as in earlier studies cited above,  a C$_{n}$ cage is modelled by  an attractive spherical potential, $U_{\rm c}(r)$, of
certain depth $U_{0}$, inner radius $r_{0}$ and thickness $\Delta$:
\begin{eqnarray}
U_{\rm c}(r)=\left\{\matrix {
-U_{0}, & \mbox{if $r_{0} \le r \le r_{0}+\Delta$} \nonumber \\
0 & \mbox{otherwise.} } \right.
\label{SWP}
\end{eqnarray}

In turn, a fullerene anion itself, C$_{n}^{-}$, is modelled as a system where the attached electron is bound into a $n\ell$ state in the field of the $U_{\rm c}(r)$ potential.

Correspondingly, the $P_{n(\epsilon)\ell}(r)$ radial wavefunctions  and $E_{n(\epsilon)\ell}$ energies of a $n\ell$ bound-state ($\epsilon\ell$ scattering-state) of the fullerene anion are
the solutions of the radial Schr\"{o}dinger equation:
\begin{eqnarray}
%\fl -\frac{1}{2}\frac{d^2P_{n(\epsilon) \ell}}{dr^2} +\left [\frac{\ell(\ell+1)}{2 r^2} +U_{\rm c}(r)\right ]P_{n(\epsilon)\ell}(r) \nonumber \\
%\fl = E_{n(\epsilon) \ell} P_{n(\epsilon)\ell}(r).
-\frac{1}{2}\frac{d^2P_{n(\epsilon) \ell}}{dr^2} +\left [\frac{\ell(\ell+1)}{2 r^2} +U_{\rm c}(r)\right ]P_{n(\epsilon)\ell}(r) \nonumber \\
 = E_{n(\epsilon) \ell} P_{n(\epsilon)\ell}(r).
\label{EqCn-}
\end{eqnarray}
The phases $\delta_{\ell}(\epsilon)$ of scattering states are determined with reference to the known  form for the radial function $P_{\epsilon\ell}(r)$
 at $r\gg 1$:
\begin{eqnarray}
P_{\epsilon\ell}(r) \approx \frac{1}{\sqrt{\pi k}}\sin\left(k r -\frac{\pi\ell}{2}+\delta_{\ell}(\epsilon)\right).
\label{P(r)}
\end{eqnarray}

Once the all needed radial functions and corresponding phases are calculated, they are plugged into equations (\ref{eqsigma})-(\ref{eqdq}) to finalize the study, i.e., to calculate
the ratio, $(\frac{d\sigma_{n\ell}}{d\Omega})/(\sigma_{n\ell}/4\pi)$, upon photodetachment of C$_{n}^{-}$.

\section{Results and Discussion}

\subsection{The $\Delta$, $r_{0}$ and $U_{0}$ parameters}

For C$_{60}$, in the literature, some inconsistency is present in choosing the magnitudes of $\Delta$, $U_{0}$ and $r_{0}$ of the model potential $U_{\rm c}(r)$.
 A better choice of the parameters with an eye on $e^{-} + {\rm C}_{60}$ elastic scattering was investigated in work \cite{U0_JPCS15}. The conclusion
was in favor of the parameters utilized in work \cite{Winstead}: $\Delta = 2.9102$ (which is twice of the covalent radius of carbon), $r_{0} = 5.262 = R_{\rm c} - (1/2)\Delta$ ($R_{\rm c} = 6.7173$ being the
known radius of the C$_{60}$ skeleton)  and $U_{0} = 7.0725$ eV (which was found by matching the known electron affinity $EA = -2.65$ eV of
C$_{60}$ with the assumption that  the $2.65$-eV-state is a ${\rm 2p}$ state). As was shown in \cite{U0_JPCS15}, the chosen
set of parameters leads to a better agreement between some of the most prominent features of $e^{-} + {\rm C}_{60}$ elastic scattering
predicted by the described model and the sophisticated \textit{ab initio} multiconfigurational Hartree-Fock approximation \cite{Winstead}.
Correspondingly, the above listed parameters will be used in the present work as well. Note, that one could opt for finding the value of $U_{0}$ by matching the electron affinity of C$_{60}$
with the assumption that the $2.65$-eV-state is a ${\rm 1s}$ state. Then, as our calculation shows, $U_{0} \approx 6.05$ eV. However, a test calculation showed that
calculated $\frac{d\sigma_{n\ell}}{d\Omega}$, obtained with the use of
$U_{0} \approx 7.072$ eV or $U_{0} \approx 6.05$ eV in the calculation, differ relatively insignificantly from each other. In the present work, as in \cite{Winstead}, the authors
choose $U_{0} \approx 7.072$ eV. Thus, for C$_{60}$, $U_{0} \approx 7.072$ eV, $r_{0} = 5.262$ and $\Delta \approx 2.91$.

For C$_{240}$,  $\Delta \approx 2.91$ (the same as for C$_{60}$, for an obvious reason), $r_{0} \approx 12.04=R_{\rm c}-(1/2)\Delta$ ($R_{\rm c}=13.46$ is the radius of the C$_{240}$ skeleton
\cite{Cabrera}). The C$_{240}$'s electron affinity $EA \approx -3.81$ eV \cite{Cabrera}. Assuming that the $3.81$-eV-state is a ${\rm 2p}$ or ${\rm 1s}$ state, one finds: $U_{0} \approx 7.92$ eV and
$U_{0} \approx 7.7$ eV, respectively. A trial calculation showed that the little difference between these two values of $U_{0}$ does not affect any noticeably calculated $\frac{d\sigma_{n\ell}}{d\Omega}$.
 In the present work, the authors choose $U_{0} \approx 7.92$ eV.

For C$_{540}$, $\Delta \approx 2.91$ (the same as for C$_{60}$ and C$_{240}$, for an obvious reason), $r_{0} \approx 18.34 =R_{\rm c}-(1/2)\Delta$ ($R_{\rm c}=19.8$ being the radius of the C$_{540}$ skeleton
\cite{C540}). Unfortunately, the authors are not aware of the value of electron affinity of C$_{540}$ needed to calculate $U_{0}$. However, it seems quite reasonable to assume that $EA$ of C$_{540}$ is not too
much different from $EA = -3.81$ eV of C$_{240}$, say, arbitrarily, $EA({\rm C_{540}}) \approx -5$ eV. The needed $U_{0}$, then, is $U_{0} \approx 9.52$ eV. Given a relative insensitivity of $\frac{d\sigma_{n\ell}}{d\Omega}$
to the variation in values of $U_{0}$, the made choice of $U_{0} \approx 9.52$ eV would hardly results in a qualitative or dramatic quantitative change of calculated $\frac{d\sigma_{n\ell}}{d\Omega}$ compared to
when the actual value of $EA(C_{540})$ would have been used in the calculation.

For C$_{1500}$, $\Delta \approx 2.91$ (the same as for C$_{60}$, C$_{240}$ and C$_{540}$, for an obvious reason), $r_{0} \approx 32 =R_{\rm c}-(1/2)\Delta$ ($R_{\rm c}=33.64$ being the mean radius of the C$_{1500}$ skeleton
\cite{GiantCns}). Unfortunately, the authors are not aware of the value of electron affinity of C$_{1500}$ needed to calculate $U_{0}$. We assign, arbitrarily, $U_{0} \approx 13$ eV, since the $U_{0}$ seems to be increasing with increasing size of a C$_{n}$ cage.

\subsection{${\rm 2p}$-photodetachment}

\begin{figure}[h]
\center{\includegraphics[width=8cm]{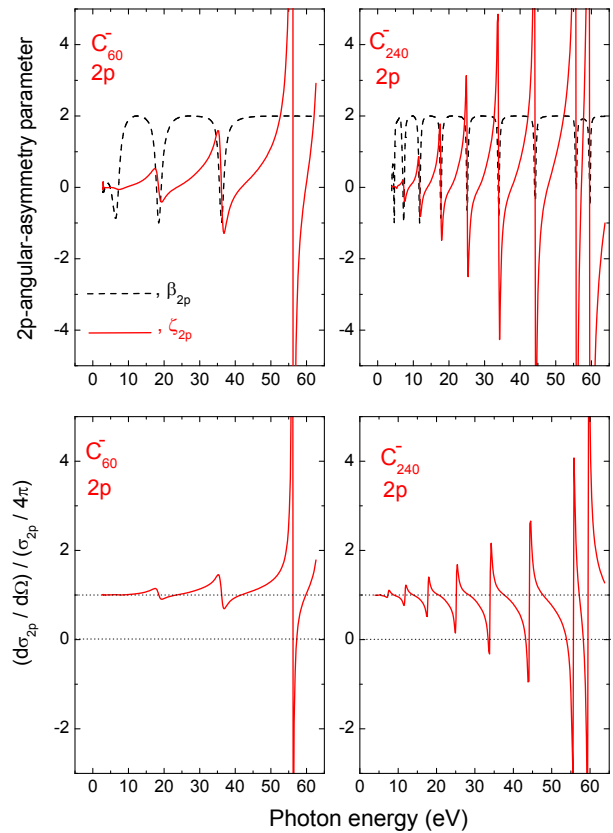}}
%\center{\includegraphics{fig1.eps}}
\caption{(Colored online) Calculated $\beta_{\rm 2p}$, $\zeta_{\rm 2p}$ and $(d\sigma_{\rm 2p}/d\Omega)/(\sigma_{\rm 2p}/4\pi)$ ratios (at the magic angle) for
${\rm 2p}$-photodetachment from the C$_{60}^{-}$ and C$_{240}^{-}$ anions.
Horizontal dotted lines are plotted for the convenience of the reader to appreciate the oscillations of the calculated ratios about unity and zero.}
\label{fig2p60240}
\end{figure}

\begin{figure}[h]
\center{\includegraphics[width=8cm]{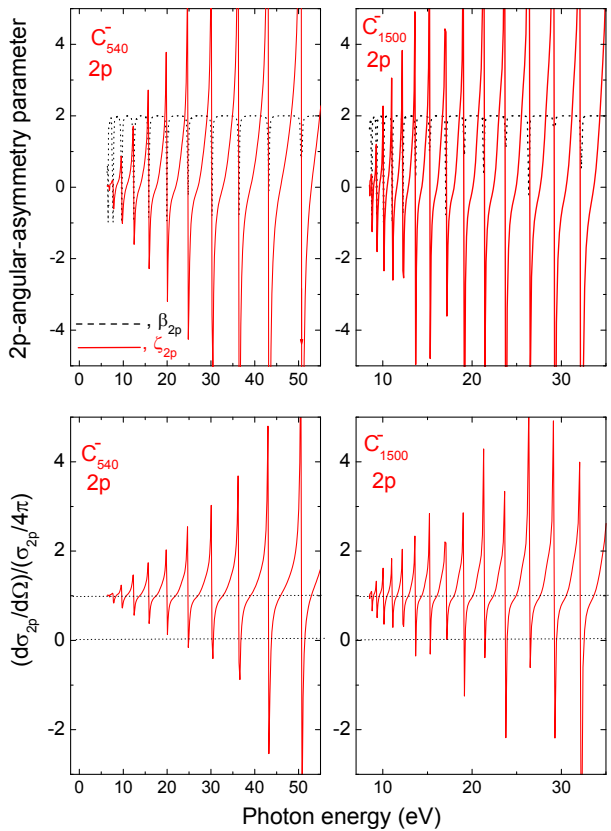}}
%\center{\includegraphics{fig1.eps}}
\caption{(Colored online) Calculated $\beta_{\rm 2p}$, $\zeta_{\rm 2p}$ and $(d\sigma_{\rm 2p}/d\Omega)/(\sigma_{\rm 2p}/4\pi)$ ratios (at the magic angle) for
${\rm 2p}$-photodetachment from the C$_{540}^{-}$ and C$_{1500}^{-}$ anions.
Horizontal dotted lines are plotted for the convenience of the reader to appreciate the oscillations of the calculated ratios about unity and zero.}
\label{fig2p5401500}
\end{figure}

We start from the discussion of the case when the attached electron resides in the ${\rm 2p}$ state of a fullerene anion.
Calculated  $\beta_{\rm 2p}$, $\zeta_{\rm 2p}$ and $(d\sigma_{\rm 2p}/d\Omega)/(\sigma_{\rm 2p}/4\pi)$ for
${\rm 2p}$-photodetachment  of C$_{60}^{-}$ and C$_{240}^{-}$ are depicted in figure~\ref{fig2p60240}, whereas those for C$_{540}^{-}$ and C$_{1500}^{-}$ - on figure~\ref{fig2p5401500}.

The calculated ratios  are seen to have the resonance character. This is by no means surprising -- the corresponding resonances are the expected confinement resonances (see the introductory section of the paper).

The striking finding of this study is that the $(d\sigma_{\rm 2p}/d\Omega)/(\sigma_{\rm 2p}/4\pi)$ ratios, and, thus, corresponding $\frac{d\sigma_{\rm 2p}}{d\Omega}$,
become negative at, and around of, certain confinement-resonance-minima. A negative $(d\sigma/d\Omega)/(\sigma/4\pi)$ does not make sense, of course, since $(d\sigma/d\Omega)$ and $\sigma$ are positive by definition. The negative value of $(d\sigma/d\Omega)/(\sigma/4\pi)$ is, therefore,
a straightforward indication of the fact that retaining only two terms -- the dipole and dipole-quadrupole terms -- in the Taylor's expansion of $(d\sigma/d\Omega)$ is an inadequate approximation, in this case. In other words, we find that, in the present case, both the $E1$-dipole and $E1$-$E2$ dipole-quadrupole approximations break down,  at corresponding energies.

Another interesting finding of this study is that the unraveled breakdown occurrences start developing at lower photon energies, and the frequency of the occurrences grows, with increasing size of C$_{n}^{-}$.

Furthermore, obviously, the $E1$ and $E1$-$E2$ breakdown occurs not only at energies, where $(d\sigma_{\rm 2p}/d\Omega)/(\sigma_{\rm 2p}/4\pi)$ is negative, but also at energies where this ratio exceeds
noticeably the unity. Indeed, when this happens, the next-order terms beyond the  $E1$-$E2$ approximation might not be negligibly small compared to both terms in the ratio, and, as such, should be accounted as well for the adequacy of the
final result for $\frac{d\sigma_{\rm 2p}}{d\Omega}$.

\subsection{$1{\rm s}$-photodetachment}

\begin{figure}[h]
\center{\includegraphics[width=8cm]{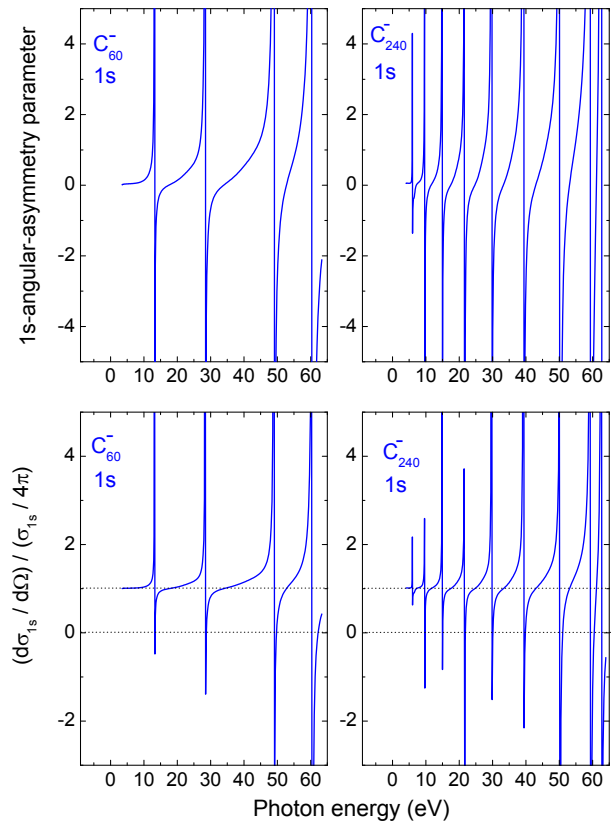}}
%\center{\includegraphics{fig1.eps}}
\caption{Calculated $\zeta_{\rm 1s}$  and $(d\sigma_{\rm 1s}/d\Omega)/(\sigma_{\rm 1s}/4\pi)$ (at the magic angle)
for ${\rm 1s}$-photodetachment of C$_{60}^{-}$ and C$_{240}^{-}$.
Horizontal dotted lines are plotted for the convenience of the reader to appreciate the oscillations of the calculated ratios about unity and zero.}
\label{fig1s60240}
\end{figure}

\begin{figure}[h]
\center{\includegraphics[width=8cm]{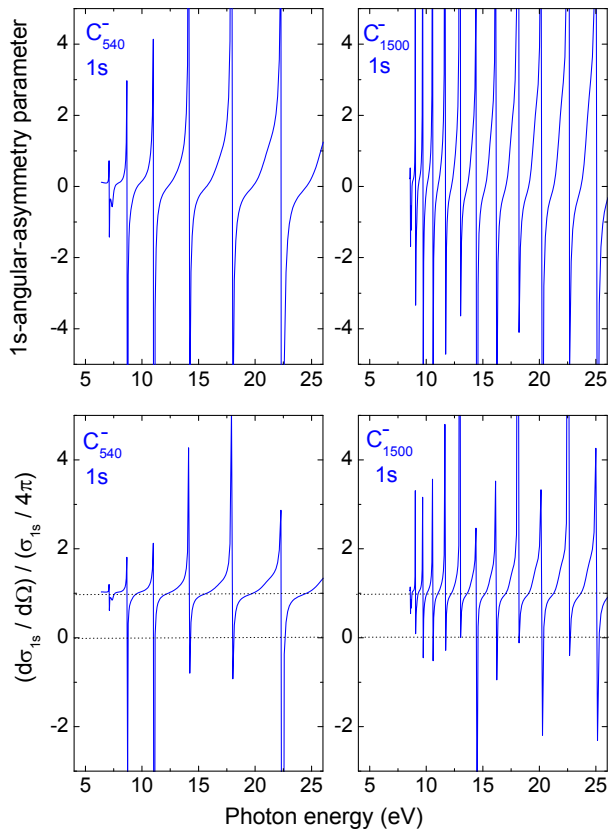}}
%\center{\includegraphics{fig1.eps}}
\caption{Calculated $\zeta_{\rm 1s}$  and $(d\sigma_{\rm 1s}/d\Omega)/(\sigma_{\rm 1s}/4\pi)$ (at the magic angle)
for ${\rm 1s}$-photodetachment of C$_{540}^{-}$ and C$_{1500}^{-}$.
Horizontal dotted lines are plotted for the convenience of the reader to appreciate the oscillations of the calculated ratios about unity and zero.}
\label{fig1s5401500}
\end{figure}

We now consider the case when the attached electron resides in the ${\rm 1s}$ state of a fullerene anion.
Calculated
$\zeta_{\rm 1s}$ and $(d\sigma_{\rm 1s}/d\Omega)/(\sigma_{\rm 1s}/4\pi)$ for ${\rm 1s}$-photodetachment of C$_{60}^{-}$ and
C$_{240}^{-}$ are depicted in figure~\ref{fig1s60240}, whereas those for C$_{540}^{-}$ and C$_{1500}^{-}$ - in figure~\ref{fig1s5401500}.
% The calculated data speak eloquently for themselves.
%

Note how, similar to ${\rm 2p}$-photodetachment, the plotted ratios become negative, i.e., the $E1$ and $E1$-$E2$ approximations break down on multiple occurrences through the photon energy range, as in the former case.

A novel finding, unveiled in the present case, is that, for a given fullerene anion,  the
 breakdown occurs at noticeably higher energies in the case of ${\rm 2p}$-photodetachment than in the case of $\rm 1s$-photodetachment, and  the
frequency of the breakdown occurrences is lower in the former than in the latter case. For the convenience of the reader,
the approximate minimum values of photon energies, at which corresponding breakdowns start occurring, are listed in table~\ref{Table1}.
\begin{table}
\caption{\label{Table1} Approximate values of the photon energy, $\omega$, where the dipole and dipole-quadrupole approximations break down for ${\rm 1s}$- and ${\rm 2p}$-photodetachment
of C$_{60}^{-}$, C$_{240}^{-}$ and C$_{540}^{-}$.}
\begin{ruledtabular}
\begin{tabular}{ddddd}
{\rm state} & {} & \multicolumn{1}{c}{$ \omega\ ({\rm eV}$)}  \\
 \cline{2-5}
      & {\rm C_{60}}  & {\rm C_{240}}  &  {\rm C_{540}} \\
 \hline
\rm 1s &13 &10 & 9\\
\rm 2p & 55 & \ 34  & \ 25 & \\
\end{tabular}
\end{ruledtabular}
\end{table}

\section{Conclusion}

 It has been found that the dipole and dipole-quadrupole approximations for photodetachment spectra of fullerene anions break down
 at a great number of the photon energies of only a few tens of eV. The frequency of the breakdown occurrences increases with increasing size of a fullerene anion. The increasing size of the anion also results in the shift of the breakdown occurrences towards the lower end of the spectrum.

 All in all, it appears that there are only relatively narrow energy bands between the confinement resonances in $\frac{d\sigma_{n\ell}}{d\Omega}$ where the dipole and dipole-quadrupole  approximations remain valid. These narrow bands are somewhat broader in the case
 of ${\rm 2p}$- than ${\rm 1s}$-photodetachment. In both cases, the bands are rapidly narrowing down with increasing size of a fullerene.

To summarize, the present study reveals that the electric dipole and dipole-quadrupole approximations alone \textit{might} be insufficient for the adequate understanding of photodetachment of fullerene anions at a great number
of photon energies as low as just few tens eV. There, terms beyond the $E1$-$E2$ approximation must, then, be accounted in corresponding calculations as well.

 In the paragraph above, we used the wording `\textit{might be}' in the first sentence, because accounting for electron correlation in the photodetachment process \textit{might} (or \textit{might not}) make the deep minima in the dipole photodetachment amplitudes to become shallower, in which case the contribution of the dipole-quadrupole correction terms (\ref{zetas}), (\ref{zetap}) in (\ref{magic}) \textit{might} (or \textit{might not}) be lessened
 to some extent. Thus, accounting for the impact of electron correlation on angle-differential photodetachment cross sections of fullerene anions is an important element to study.
Such study,
however, as was pointed out in Introduction, is beyond the scope of the present work. Rather, we believe that the results
of the present study will encourage other researchers to
perform experimental or more sophisticated theoretical
studies of angle-differential photodetachment cross
sections of fullerene anions, now that the present work has
highlighted the most intriguing parts of the problem that are in need of a deeper study.

\section*{References}

\end{document}